\documentclass[twoside,english]{elsarticle}
\usepackage[T1]{fontenc}
\usepackage[utf8]{inputenc}
\usepackage{float}
\usepackage{textcomp}
\usepackage{amsmath}
\usepackage{ amssymb }
\usepackage{graphicx}
\usepackage{xcolor}

\makeatletter

\providecommand{\tabularnewline}{\\}

\journal{Journal of Magnetism and Magnetic Materials}

\usepackage[version=3]{mhchem}

\makeatother

\usepackage{babel}
\begin{document}

\begin{frontmatter}{}

\title{Magnetization reversal and anisotropies in buffered transition-metal alloys thin films}

\author[label1]{A. Lo Giudice}
\author[label1,label2]{A. Roman}
\author[label1]{L.B. Steren}
\affiliation[label1]{Instituto de Nanociencia y Nanotecnologia CNEA CONICET - Nodo Constituyentes, 1650 San Martin,  Buenos Aires, Argentina}
\affiliation[label2]{Dpto. Micro y Nanotecnologia, Centro Atomico Constituyentes CNEA, 1650 San Martin, Buenos Aires, Argentina}

\begin{abstract}

Interest in planar Hall effect (PHE) sensors has re-emerged in recent years due to their promising potential for a wide range of applications, particularly in biotechnology. Sensor sensitivity can be enhanced by lowering the effective anisotropy field; however, this favors magnetic domain formation during magnetization reversal, leading to hysteretic responses. Therefore, precise control of magnetic anisotropy and magnetization reversal is essential to balance sensitivity and stability in PHE sensors. In this work, we investigate the magnetic anisotropy and magnetization reversal mechanisms of Ni–Fe- and Co–Fe-based multilayers grown on various metallic buffer layers and deposited with and without an external magnetic field, in order to evaluate the effects of the buffer layers and field-assisted deposition on the resulting magnetic anisotropy. NiFe films exhibit a dominant uniaxial anisotropy mainly determined by the applied field during growth, with an anisotropy constant of approximately $3 \frac{kerg}{cm^3}$, largely independent of the buffer layer. In contrast, the magnetic anisotropy of CoFe films is dominated by the buffer layer, resulting in a biaxial magnetic response. In particular, Ag-buffered films deposited under an external magnetic field exhibit a biaxial anisotropy with values up to $14.88 \frac{kerg}{cm^3}$. The magnetization reversal mechanism of each system was deduced from the analysis of the coercive field angular dependence.

\end{abstract}

\end{frontmatter}{}

\section{Introduction}

Planar Hall Effect (PHE) sensors have regained interest from the magnetic sensor community due to their high low field sensitivity, low temperature drift, high signal-to-noise ratio, low cost and ease of fabrication \citep{SCHUHL,lim2022advances}. Different applications motivated the search for high thermal stability and low power consumption sensors \citep{Freitas, Thanh, Granell}. PHE sensors are typically based on soft magnetic thin films due to their low coercive field and high anisotropic magnetoresistance (AMR) ratio \citep{Lenz}. Tuning the properties of magnetic thin films to satisfy the requirements of PHE sensors and achieve optimal responses is the most important challenge faced nowadays. 

The Planar Hall Effect  and anisotropic magnetoresistance (AMR) are related phenomena, arising from intrinsic spin-orbit coupling within ferromagnetic materials. A key feature of the PHE is its linear response to magnetic field around zero field. A significant itation of the PHE is its hysteretic behavior, primarily caused by nonuniform magnetization across the sensor element, which complicates absolute measurements of the magnetic field. The maximum field for generating a useful signal is determined by the coercive field. The sensors' sensitivity is enhanced by lowering the effective anisotropy field but reducing the anisotropy field can lead to the formation of magnetic domains in the sensing layer, resulting in non-linear field response and broader hysteresis. In contrast, larger anisotropy fields stabilize the sensing layer in a nearly single-domain state, minimizing hysteresis and nonlinearities.  Consequently, finding the right balance of anisotropies and magnetization reversal mechanisms remains the key challenge in the design and optimization of PHE-based sensors \cite{elzwawy2021current, lim2022advances}.

Nickel-iron alloys, particularly Ni$_{80}$Fe$_{20}$ (NiFe, Permalloy), are ideal ferromagnetic materials to use in PHE sensors due to their strong spin-orbit interaction and high Curie temperature. NiFe films typically show AMR values of 2\%-2.2\% at low magnetic fields that can be increases up to 3.5\%  with tailored deposition conditions. By optimizing parameters such as substrate temperature, buffer layer thickness, and deposition field, NiFe-based sensors achieve ultra-low magnetic field detection down to 5 pT, as demonstrated in recent studies
\cite{elzwawy2021current, schuhl1995low, wang2013studies}.

The Co$_{90}$Fe$_{10}$ (CoFe) alloy is also a promising candidate for spintronic applications due to its magnetic properties, including large saturation magnetization, high Curie temperature, and high permeability. These properties enable CoFe films to deliver a reliable performance, even under challenging conditions. Magneto-transport experiments performed on Co$_x$Fe$_{1-x}$ single-crystal samples have reported AMR ratios as high as 2.4\%.  The magnetoresistance was found to be strongly dependent on the crystalline orientation, with values ranging from 0.2\% to 2.1\% depending on the direction in which the effect was measured \cite{su2023investigating,zeng2020intrinsic, zeng2020role}. It must be noted that the magnetic anisotropy of this alloy is much larger than that of permalloy so higher coercivities are expected for this system. \cite{Ingvarsson}

To achieve an optimal magnetic material  for PHE sensors, various strategies were explored \citep{Vayalil}. It is well known that using underlayers can significantly alter the magnetic properties of thin films \citep{Jeong}. Furthermore, tuning the coercivity of magnetic films can be reached by magnetic annealing or deposition under applied magnetic fields \citep{Saravanan}.
The aim of this paper is to analyze the magnetic anisotropy and magnetization reversal mechanisms of magnetic structures designed for PHE sensors.  

Here, the effects of various metallic buffer layers as well as the influence of an external magnetic field applied during deposition on multilayers based on NiFe and CoFe alloys are investigated and discussed. 

The choice of the buffer layers was driven by the need to enhance the sensitivity of planar Hall effect (PHE) sensors, which requires ferromagnetic films with low coercive fields and well-controlled magnetic anisotropy \cite{CoFe_different_underlayers}. Tantalum is widely employed as a buffer layer for permalloy films. When deposited as a thin film, Ta typically exhibits poor crystallinity, a property related to its high melting point relative to the deposition temperature \cite{MRSonline}. More important, the poor structural order of Ta does not impose its own crystalline texture onto the ferromagnetic layer; instead, Ta acts as an effective diffusion barrier and prevents the formation of an amorphous interfacial layer induced by the native \ce{SiO_2} \cite{MRSonline,JMMM95}. This behavior favors the development of a well-defined (111) texture in permalloy films, which has been correlated with reduced coercivity \cite{JAll2013}. In contrast, more crystalline buffers such as Cr can interfere with the ferromagnetic growth, leading to a broader distribution of anisotropy axes and   an increase of the coercive field \cite{MRSonline}. An analogous approach motivates the use of W buffer layers, whose high melting point and thermal stability effectively suppress interdiffusion at the interface and reduce defect formation that may otherwise enhance coercivity \cite{JMMM95, JMMM93}. Cu and Ag buffers were selected due to their extensive use in the fabrication of epitaxial or highly ordered metallic films, including Fe-, Co-, Ni-, and CoFe-based systems \cite{SciRep17,IEEE05,ASS05}. Although epitaxial growth is not expected under our deposition conditions (room-temperature sputtering), both Cu and Ag are known to promote strain relaxation and enhanced structural order compared to direct growth on Si or \ce{SiO_2} \cite{SciRep17}. Such improvements in crystalline order can directly affect magnetic anisotropy, grain size distribution, and magnetization reversal processes, which are closely linked to coercivity and Barkhausen noise \cite{JAP97}. Overall, the selected buffer layers allow us to explore two complementary approaches for coercivity reduction: diffusion control using high-melting-point buffers, and enhanced crystalline order using buffers favorable to epitaxial growth.

\section{Materials and Methods}

The trilayers M/FM/M (M: Ta, Ag, Cu, W; FM: NiFe, CoFe)
used for this study were grown on clean
Si(100) substrates by DC magnetron sputtering using a commercial AJA ATC Orion 8 equipped with 4 sputtering cannons. The materials selected as the underlayers, M, were
also used as capping layers to avoid film oxidation. The thickness of the M layers was kept fixed at
5nm along the whole set of samples. The thickness of the ferromagnetic layers was $t_{FM}$ = 20 nm for all systems. The sputtering deposition parameters for each material are detailed in Table \ref{tab:Sputtering-deposition-parameters}.
\begin{table}[htbp]

\centering{}%
\begin{tabular}{|c|c|c|c|}
\hline 
Material & Ar Pressure (mTorr) & Sputtering Power (W) \tabularnewline
\hline 
Ta & 6 & 26 \tabularnewline
\hline 
Ag & 2 & 25\tabularnewline
\hline 
Cu & 3 & 51\tabularnewline
\hline 
W & 3 & 30 \tabularnewline
\hline 
NiFe & 3 & 100\tabularnewline
\hline 
CoFe & 3 & 76\tabularnewline
\hline 
\end{tabular}
\caption{\textsc{\label{tab:Sputtering-deposition-parameters}Sputtering deposition
parameters}}
\end{table}

In order to analyze the effect of a magnetic field, $H_d$, during deposition two complementary series of multilayers were grown under a $H_d=300$ Oe applied parallel to the film\textquoteright s plane. $H_d$ was produced by an array of permanent Neodymium-Iron-Boron magnets installed in a custom substrate holder. The samples were named, throughout this article, as FM\_M (NH/H), whether the trilayer was deposited under a magnetic field (H) or not (NH). 
The deposition rate of all deposited films -with and without an applied magnetic field during deposition- were periodically calibrated using X Ray Reflectometry and Atomic Force Microscopy measurements.

The crystalline structure \textcolor{blue}{was analyzed by Grazing-Incidence-X-Ray Diffraction (GIXRD) using Cu-K$_\alpha$ radiation}. Surface roughness and grain sizes of the samples were probed Atomic Force Microscopy (AFM) and Scanning Electron Microscopy (SEM).
The magnetic properties of the trilayers were studied by magnetization loops measured at room temperature using a Lake Shore Vibrating Sample Magnetometer. The magnetic anisotropy of the different systems was derived from the in-plane angular dependence of the magnetization curves.

\section{Results} 
\subsection{Crystalline texture and morphology}
\begin{figure}
\centering{}\includegraphics[width=0.6\textwidth]{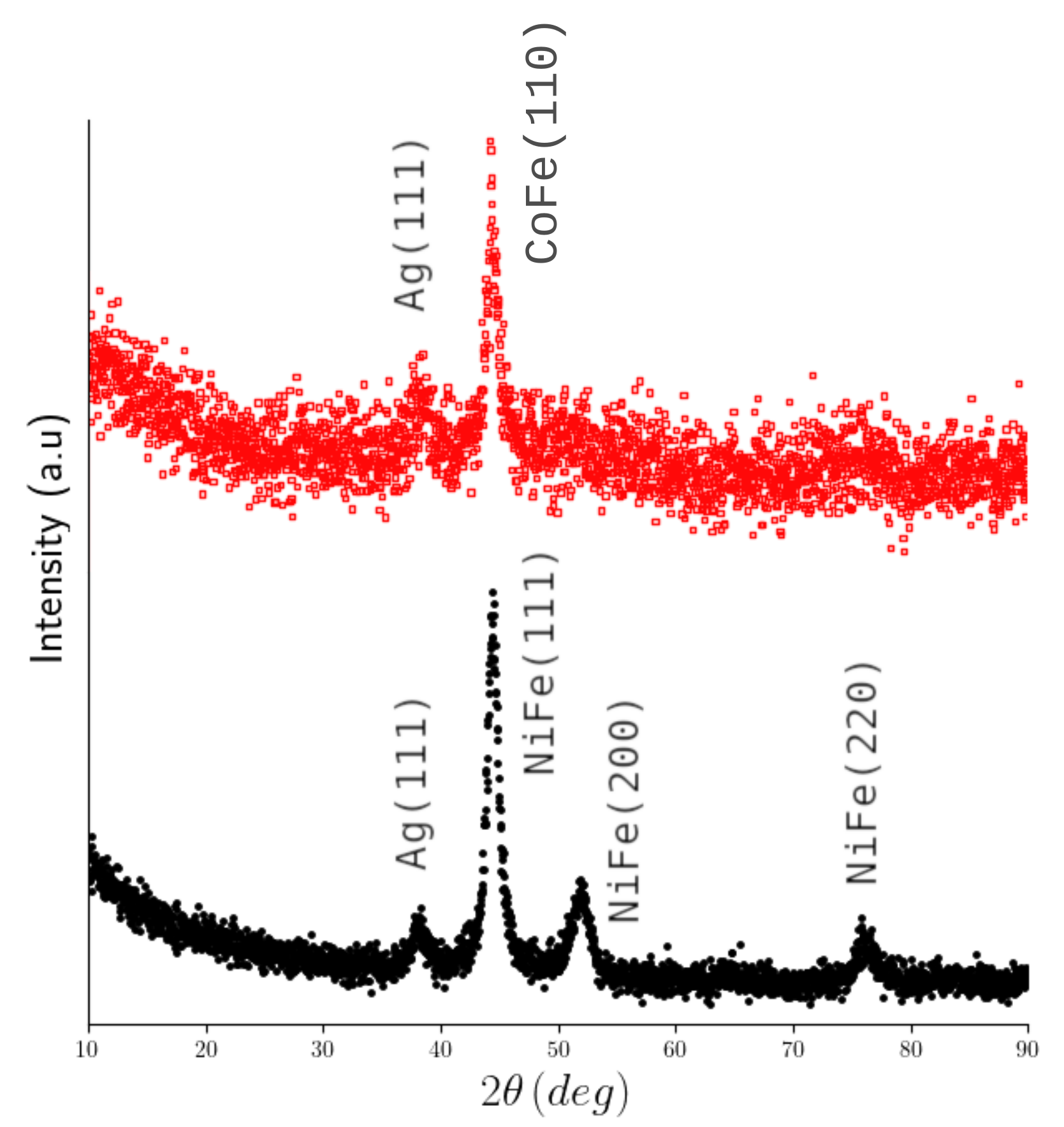}
\caption{\label{fig:GIXRD}GIXRD patterns of 100nm-thick CoFe  and
NiFe films with Ag buffer and capping.}
\end{figure}

GIXRD patterns of 100 nm thick NiFe and CoFe films, buffered and capped with Ag, are shown in Fig. \ref{fig:GIXRD}. The NiFe diffractogram exhibits multiple peaks indicative of a polycrystalline structure, which were indexed to the fcc phase. As expected for this structure, the (111) reflection is the most intense compared to the other reflections. The position of this peak allowed us to estimate a lattice parameter of 3.53 \AA, in agreement with previous references  (\cite{Yeh, guittoum}). Similarly, the GIXRD pattern of CoFe indicated a polycrystalline bcc structure, with a lattice parameter of 2.84 \AA, calculated from the (110) reflection - the most intense peak in this structure. Unlike the bulk alloy which has a fcc structure \cite{finkel}, CoFe thin films were usually found with a bcc structure whose lattice parameter varies between 2.86\AA \cite{alper} and 2.88\AA \cite{cakmaktepe} in coincidence with our results.   Using the Scherrer formula \cite{scherrer1918bestimmung}, the grain size corresponding to the (111) and (110) peaks was estimated to be $(17\pm1)$ nm for both materials.

\begin{figure}
\centering{}
\includegraphics[width=0.9\textwidth]{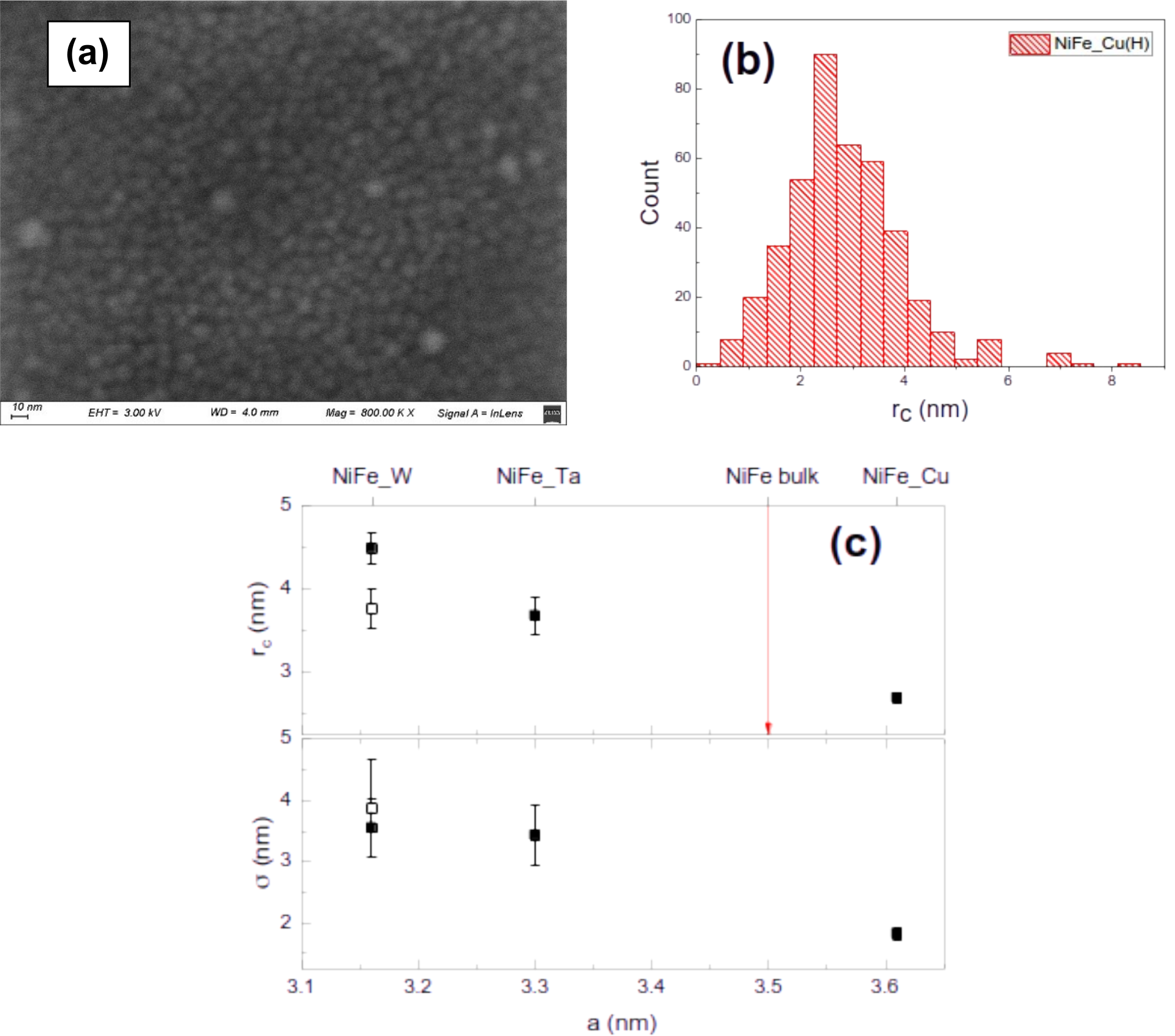}
\caption{\label{fig:SEM}{\textcolor{blue}({a) SEM image and (b) grain size distribution of a
NiFe\_Cu(H) bilayer; (c) Gaussian fit parameters, mean grain radius $r_c$ and variance
$\sigma$, for samples grown under magnetic field (full symbols) and without (open
symbols)}}}
\end{figure}

A quantitative analysis of the grain morphology of NiFe films was performed for all buffer layers by extracting grain size distributions from SEM images (Fig. \ref{fig:SEM}(a)) using the Gwyddion software. The corresponding grain size histogram of the NiFe\_Cu(H) bilayer is representatively shown in Fig. \ref{fig:SEM}(b) and a summary of the extracted Gaussian fit parameters (mean grain size \& variance) for all the bilayers is shown in Fig. \ref{fig:SEM}(c). A variation in grain size is observed depending on the buffer layer, decreasing from approximately 4 nm for samples deposited on W to about 2 nm for those deposited on Cu. A clear dependence of the grain size distribution on the buffer layer is observed: in particular, films deposited on W exhibit a significantly broader grain size distribution, whereas films grown on Cu buffers show a much narrower distribution, indicative of a more homogeneous grain structure. This behavior can be correlated with the melting point of the buffer layers. W has a very high melting point (3422 °C), while Cu presents a considerably lower one (1084.6 °C), with Ta lying in between (3017 °C). As reported in previous works \cite{MRSonline}, the mobility of adatoms during the deposition process depends on the difference between the melting point of the underlying layer and the substrate temperature. A higher melting point relative to the substrate temperature leads to reduced adatom mobility, resulting in increased structural disorder in the deposited film.
We also compared the grain size distributions of films deposited with and without an applied magnetic field. No noticeable differences were observed for Ta- and Cu-buffered samples. In contrast, as we show in Fig. \ref{fig:SEM}(c), W-buffered films exhibit clear differences: films deposited under an applied magnetic field show a larger average grain size and a narrower grain size distribution, indicating a more ordered grain structure. This effect can be attributed to the magnetic field promoting a preferential alignment of Fe ions during growth \cite{kneer1966origin}, thereby enhancing structural order. While a similar mechanism may be present in all systems, it becomes experimentally observable only for W-buffered films, where the initially broader grain size and distribution allow the field-induced ordering to be more clearly resolved.
 
\subsection{Magnetic properties}\label{analisis_remanencia}
\subsubsection{Experiments}
\begin{figure}
\begin{centering}
\includegraphics[width=0.7\textwidth]{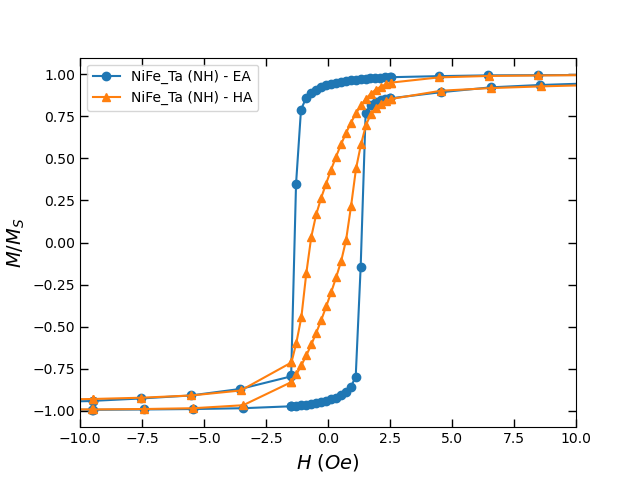}

\caption{\label{fig:NiFe-ciclos} In-plane magnetization loops of a NiFe\_Ta (H) structure measured at T=300K,  along both the uniaxial easy ($\color{blue}\bullet$, EA) and hard ($\color{orange}\blacktriangle$, HA) axes respectively.}
\end{centering}
\end{figure}

Magnetization loops were measured with the magnetic field oriented in the plane of the films at azimuthal angles varying from $0^{\circ}$ to $360^{\circ}$. As expected, all the samples present an easy-plane anisotropy associated to shape and most of them, in addition,  exhibit a uniaxial anisotropy (Fig. \ref{fig:NiFe-ciclos}). Changes observed in the shape of the ferromagnetic hysteresis loops as buffer layers are inserted between substrates and magnetic layers suggest that they are affecting the magnetization reversal mechanism of the structures.

The in-plane angular dependence (Figs. \ref{fig:NiFe_mr_hc} and \ref{fig:CoFe-mr-hc}) of the remanent magnetization, $M_r$,  and coercivity, $H_c$, were analyzed in detail to get a deeper understanding of the magnetization reversal mechanisms and determine the magnetic anisotropies of the samples. The remanent magnetization depends mainly on the sample's magnetic anisotropies while the coercivity is mostly related to the reversal mechanism, such as coherent rotation or domain-wall motion. 

\subsubsection{Models} 
\paragraph{Remanent magnetization} 
The analysis of the remanent magnetization was done assuming that the samples' volume is distributed in three fractions based on their magnetic anisotropy. The remanent magnetization is, thus, described by: 

\begin{equation}
\label{eq:remanencia_modelo}
    m_r=x_1\cdot \ m_{r1}(\varphi_H)\,+\,x_2\cdot\ m_{r2}(\varphi_H,K)\,+\,x_3 \cdot 0.5
\end{equation}

$x_1$ represents the fraction of grains that exhibit a single uniaxial anisotropy direction. The corresponding angular dependence of the remanent magnetization, follows the classical Stoner–Wohlfarth model \cite{stoner1948mechanism}.  $x_2$   accounts for the fraction of grains that exhibit superposed biaxial and uniaxial anisotropies. The remanence angular dependence was obtained by minimizing the magnetic free energy of the system. Finally, $x_3$  represents the fraction of grains with randomly distributed easy-axis orientations. In the Stoner–Wohlfarth framework, a completely random distribution of uniaxial anisotropy axes leads to an average remanent magnetization equal to 0.5 \cite{stoner1948mechanism}. The coexistence of regions with different types of magnetic anisotropy and their description as weighted contributions to the total magnetic response has been previously adopted in the literature \cite{belyaev2016competing}.

The term $m_{r1}(\varphi_H)$ is deduced from the single-domain uniaxial anisotropy Stoner–Wohlfarth model \cite{stoner1948mechanism}, and is given by:
\begin{equation}
m_{r1}=\frac{M_{r1}}{M_s}=\cos(\varphi_H).
\end{equation}

\begin{figure}[H]
\begin{centering}  
    \includegraphics[width=0.6\textwidth]{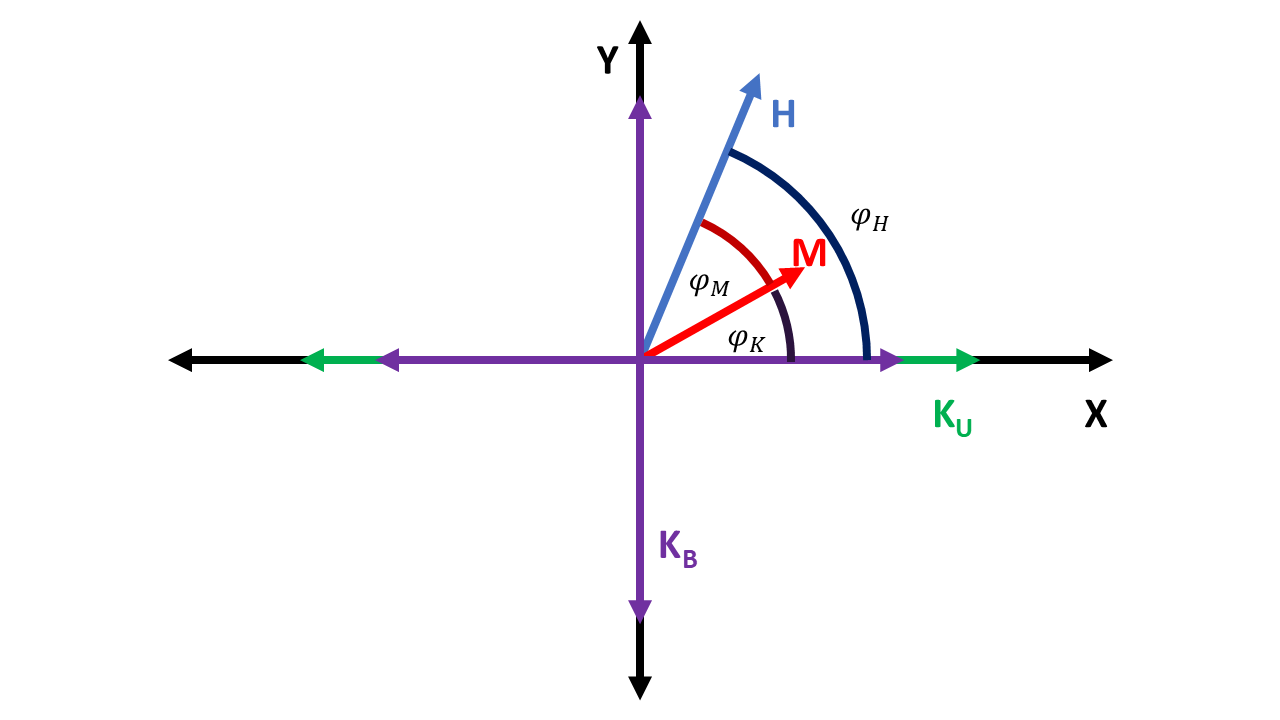}
    \caption{\label{fig:esquema-KU_KB}Schematic representation of the magnetization and magnetic field vectors in the description of the free energy with respect to the anisotropy axes. }
\end{centering}
\end{figure}

The $m_{r2}(\varphi_H,K)$ contribution is deduced assuming a Stoner coherent rotation of a single-domain volume \cite{stoner1948mechanism} with two different anisotropies, uniaxial and biaxial. Within this model, the two-fold anisotropy axis is oriented along one of the four-fold anisotropy axes, x-axis (Fig. \ref{fig:esquema-KU_KB}). The free energy density is given by:

\begin{equation}
\label{eq:free_energy_KU_KB}
    E=K_{U}\sin^{2}(\varphi_K)+\frac{K_B}{4}\sin^{2}(2\varphi_K)-H\cdot M_s\cos(\varphi_M)
\end{equation}

where $K_U$ and $K_B$ are the uniaxial and biaxial anisotropy constants, $H$ is the applied magnetic field, $M_S$ the saturation magnetization, $\varphi_K$ the angle between the magnetization and the uniaxial axis of anisotropy, and $\varphi_M$ the angle between the applied magnetic field and the magnetization vector (Fig. \ref{fig:esquema-KU_KB}). $m_{r2}(\varphi_H,K)$  is derived  by minimizing the energy density (Eqn.\ref{eq:free_energy_KU_KB}). Analytical solutions were found to depend on the anisotropies' ratio $K=\frac{K_B}{K_U}$. For $0\leq K<1$ the remanence takes the form:
 \begin{equation}
    m_{r1}=\frac{M_{r1}}{M_s}=cos(\varphi_H)
 \end{equation}

while for $K\geq 1$ the result is
 \begin{equation}
    m_{r2}=\frac{M_{r2}}{M_s}=\left\{\begin{array}{lcc} 
    cos(\varphi_H)&for& |sin(\varphi_H)|<\sqrt{(\frac{1}{2}+\frac{1}{2K})}\\
    sin(\varphi_H)&for& |sin(\varphi_H)|\geq\sqrt{(\frac{1}{2}+\frac{1}{2K})}
    \end{array}\right.
    \label{biaxial+uniaxial_mr}
 \end{equation}

Thus, for $0\leq K < 1$,  the resultant remanence $m_{r2}(\varphi_H,K)$ is aligned along the uniaxial anisotropy axis direction.  Instead, for $K\geq1$ there is a critical angle, determined by the relationship between the anisotropies, at which the remanent magnetization shifts from the $x$ to the $y$-axis. 

\paragraph{Coercivity} 
While the remanent magnetization provides information about the different anisotropies present in the trilayers, the analysis of the angular dependence of the coercive field  gives a valuable insight into the mechanisms of magnetization reversal. The Stoner-Wohlfarth model describes the change of magnetization state by coherent rotation of single-domain particles with uniaxial anisotropy \citep{SW_ADC}. According to this model, the coercive field in the easy-axis direction, $H^{EA}_{C}$  is equal to the saturation field $H^{HA}_{sat}$ measured along the hard axis and is given by 

\begin{equation} 
H^{EA}_{c}= H^{HA}_{sat} = \frac{2K_u}{M_s},
\label{eq:Stoner-Wohlfarth model_2} 
\end{equation}

The coercive field's angular dependence for a single domain with uniaxial and biaxial anisotropies can be determined by minimizing the free energy (Eqn. \ref{eq:free_energy_KU_KB}) numerically. Differing from the previous case,  $H^{EA}_{C}$   turns out to be: 

\begin{equation}
    H^{EA}_{C}=\frac{2K_u(K+1)}{M_s}
    \label{Hc_uni+biax}
\end{equation}

 In our study it is generally observed that $H^{EA}_{C} < H^{HA}_{sat}$. These results indicate that the reversal process occurs through either domain wall movement and/or coherent rotation. Suponev et al \citep{Suponev} proposed a theory for multi-domain systems, in which it is assumed that there are two types of magnetic domains, or two magnetic phases. The angular dependence of the coercive field is,  for this case:

\begin{equation}
H_{c}^{TP}=H^{EA}_{C}\cdot\:\left[\frac{\cos(\varphi_H)}{\frac{1}{\gamma}\sin^{2}\varphi_H,+\cos^{2}\varphi_H}\right],\quad\gamma=\frac{N_{A}+N_{X}}{N_{Y}}\label{eq:two-phase}
\end{equation}

where $N_{X}$ and $N_{Y}$ are the demagnetizing factors of an ellipsoid along its principal axes, while $N_{A}$ takes into account the contribution of other anisotropies beyond shape anisotropy. The magnetization reversal mechanisms are defined by the value of $\gamma$. Experimental results, previously reported, exhibiting the angular dependence described by Eqn. \ref{eq:two-phase}  have shown that for angles close to the hard axis the magnetization reversal occurs primarily via domain rotation ($\gamma < 2$). In contrast, at angles close to the easy-axis the reversal of magnetization takes place  also by domain-wall movement ($\gamma \geq 2$) \citep{roman2024magnetization}.  Furthermore, as $\gamma$ increases, the expression becomes similar to the Kondorsky model \citep{kondorsky1940hysteresis}, which describes magnetization reversal solely through domain wall movement. Conversely, decreasing $\gamma$ broadens the angular range where reversal occurs via domain rotation.

\begin{figure*}[htp]
\begin{centering}
\includegraphics[width=1.2\textwidth]{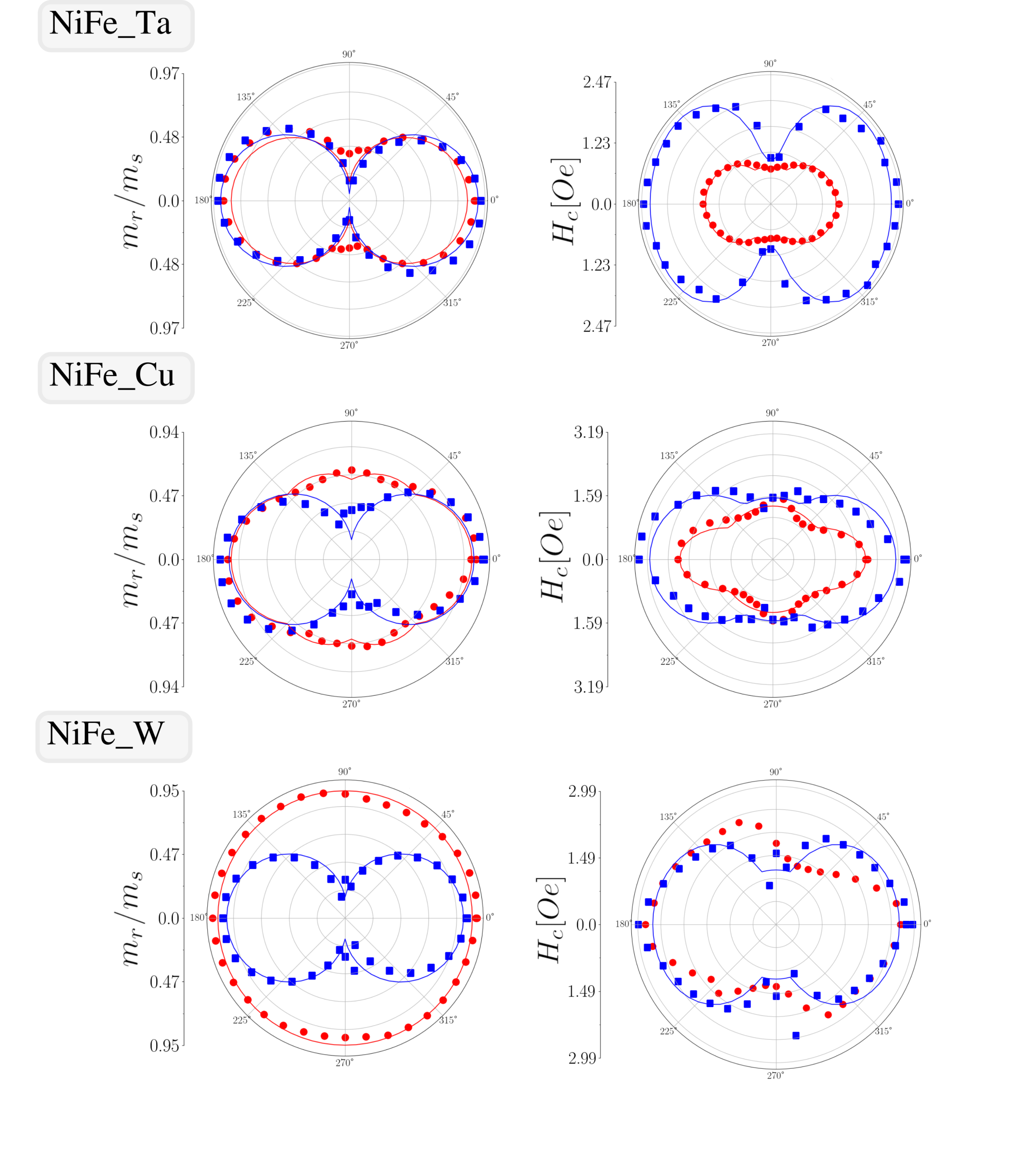} 
\end{centering}
\caption{\label{fig:NiFe_mr_hc}Polar plots of the normalized remanent magnetization (left) and coercive field (right) for NiFe-based films deposited with (\textcolor{blue}{$\blacksquare$}) and without (\textcolor{red}{$\bullet$}) an applied magnetic field. The solid lines are fits to $m_r/m_s$ using Eqn.\ref{eq:remanencia_modelo} and to $H_c$ using Eqn. \ref{eq:two-phase}, respectively.}
\end{figure*}

The angular dependence of the remanent magnetization of NiFe\_M samples is shown in Fig. \ref{fig:NiFe_mr_hc}(left).  This data was fitted by Eqn. \ref{eq:remanencia_modelo} which describes the superposition of single-domain grains with biaxial and uniaxial anisotropies.

The parameters derived from the curves fitting are presented in Table \ref{tabla:NiFe_mr}. 
These results show that the magnetic anisotropy of the NiFe\_M(NH) multilayers is strongly influenced by the buffers.
 On one hand, tantalum buffered samples present in-plane uniaxial anisotropy while on the other hand,  NiFe\_W(NH) structures were found to be in-plane magnetically isotropic. 
Copper buffered samples, instead, show a bigger biaxial anisotropy compared to the uniaxial one.  

\begin{table}[ht]
\centering{}
\resizebox{\textwidth}{!}{
\begin{tabular}{|l|cccccccc|}
\hline
       \multicolumn{1}{|c|}{Sample}
       & \multicolumn{1}{c|}{$x_1$}    & \multicolumn{1}{c|}{$x_2$}   & \multicolumn{1}{c|}{$x_3$}    & \multicolumn{1}{c|}{$H^{EA}_{C}$ (Oe) }&\multicolumn{1}{c|}{$H^{HA}_{sat}\ (Oe) $}&\multicolumn{1}{c|}{$K_U\ (\mathrm{\frac{kerg}{cm^3}}) $}&\multicolumn{1}{c|}{$K_B\ (\mathrm{\frac{kerg}{cm^3}}) $}& \multicolumn{1}{c|}{$\gamma$}\\ \hline
NiFe\_Ta(NH) & \multicolumn{1}{c|}{0.74} & \multicolumn{1}{c|}{0}    & \multicolumn{1}{c|}{0.26}  &\multicolumn{1}{c|}{1.31} &\multicolumn{1}{c|}{3.81}   &\multicolumn{1}{c|}{1.52} &\multicolumn{1}{c|}{-}  &\multicolumn{1}{c|}{1.41}   \\ \hline
NiFe\_Cu(NH) & \multicolumn{1}{c|}{0.28}  & \multicolumn{1}{c|}{0.42} & \multicolumn{1}{c|}{0.3}  &\multicolumn{1}{c|}{2.52}&\multicolumn{1}{c|}{3.64}  &\multicolumn{1}{c|}{0.06} &\multicolumn{1}{c|}{0.95} &\multicolumn{1}{c|}{-} \\ \hline
NiFe\_W(NH) & \multicolumn{3}{c|}{isotropic}&\multicolumn{1}{c|}{2.80}&\multicolumn{1}{c|}{2.00}&\multicolumn{1}{c|}{-}&\multicolumn{1}{c|}{-}&\multicolumn{1}{c|}{-}  \\ \hline
NiFe\_Ta(H)  & \multicolumn{1}{c|}{0.90} & \multicolumn{1}{c|}{0}    & \multicolumn{1}{c|}{0.10} &\multicolumn{1}{c|}{2.46}&\multicolumn{1}{c|}{7.61} &\multicolumn{1}{c|}{3.04}&\multicolumn{1}{c|}{-}&\multicolumn{1}{c|}{2.65}          \\ \hline
NiFe\_Cu(H)  & \multicolumn{1}{c|}{0.74} & \multicolumn{1}{c|}{0}    & \multicolumn{1}{c|}{0.26}  &\multicolumn{1}{c|}{2.87}  &\multicolumn{1}{c|}{7.94}&\multicolumn{1}{c|}{3.17} &\multicolumn{1}{c|}{-}  &\multicolumn{1}{c|}{0.93}    \\ \hline
NiFe\_W(H)  & \multicolumn{1}{c|}{0.70} & \multicolumn{1}{c|}{0}    & \multicolumn{1}{c|}{0.30} &\multicolumn{1}{c|}{2.39}&\multicolumn{1}{c|}{9.01} &\multicolumn{1}{c|}{3.60}&\multicolumn{1}{c|}{-} &\multicolumn{1}{c|}{0.99}     \\ \hline
\end{tabular}
}
\caption{\label{tabla:NiFe_mr}Parameters extracted from fitting the angular dependence of remanent magnetization and coercive field for the Ni-Fe trilayers. The saturation field ($H^{HA}_{sat}$) and uniaxial anisotropy constants ($K_U$ and $K_B$) were obtained from hysteresis loops measured with the magnetic field applied along the hard and easy axes for the same set of samples.}
\end{table}

In the NiFe\_M(H) samples, on the contrary, the magnetic field applied during deposition induces a notable enhancement of the uniaxial anisotropy, becoming  the dominant anisotropy over the others.  These results are thus better described by a predominant uniaxial anisotropic region that coexists with smaller zones with randomly distributed easy-axes. In films buffered with Ta, the fraction of the samples with uniaxial anisotropy increases from 70\% to 90\% while on Cu buffered structures the biaxial anisotropy is completely blurred out and regions with uniaxial anisotropy increase from 28\% to 74\%. Finally, tungsten buffered films acquire a uniaxial anisotropy of the same order of magnitude found in the other systems. 

The angular dependence of the coercivity (Fig. \ref{fig:NiFe_mr_hc} , right) in NiFe\_M  samples with dominant uniaxial anisotropy is described by the two-phase model. In the case of NiFe\_Cu(NH) samples where the biaxial anisotropy is dominant, $Hc(\psi_H)$ indicates that the magnetization reverses by coherent rotation. In NiFe\_Ta structures, two distinct reversal mechanisms were identified. The sample deposited without an applied magnetic field exhibits $\gamma < 2$, consistent with domain rotation, whereas the sample deposited under a magnetic field exhibits $\gamma > 2$, indicative of domain wall motion. For other samples grown under field, $\gamma < 2$, which indicates magnetization reversal predominantly by rotation. The non-zero values of the coercive field measured along the hard axis indicates the presence of small regions with randomly distributed easy axes, also observed in the remanent magnetization measurements.

\begin{figure*}[tp]
\begin{centering}
\includegraphics[width=1.2\textwidth]{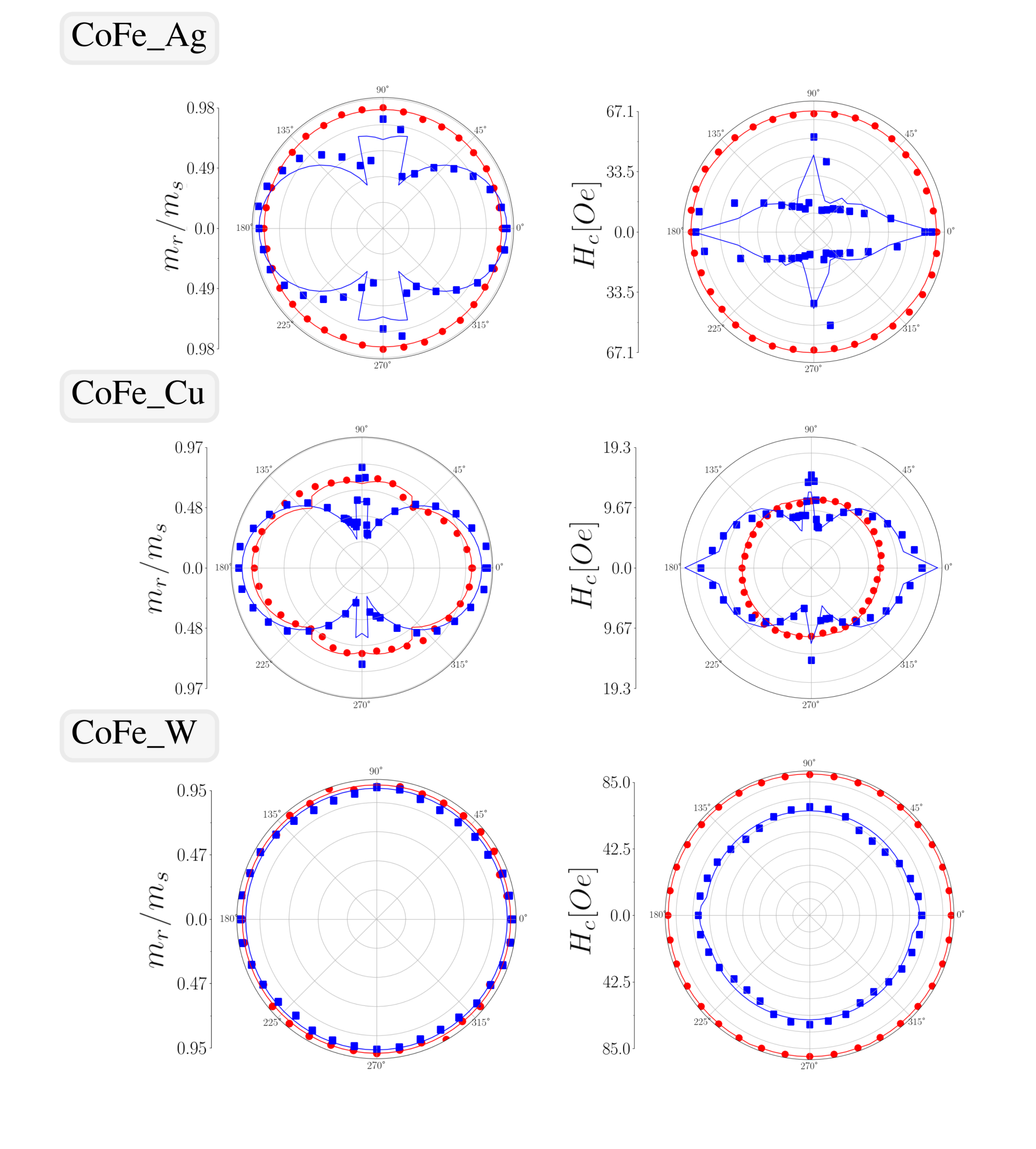} 
\end{centering}
\caption{\label{fig:CoFe-mr-hc}Polar plots of the normalized remanent magnetization (left) and coercive field (right) for CoFe-based films deposited with (\textcolor{blue}{$\blacksquare$}) and without (\textcolor{red}{$\bullet$}) an applied magnetic field. The solid lines are fits to $m_r/m_s$ using Eqn.\ref{eq:remanencia_modelo} and to $H_c$ using Eqn. \ref{eq:two-phase}, respectively.}
\end{figure*}

The effects of buffer layers and an applied magnetic field during deposition on the magnetic properties of CoFe films are shown in Fig. \ref{fig:CoFe-mr-hc} and Table \ref{tabla:CoFe_mr}. As expected, the effect of buffers is important. The copper buffer induces a biaxial anisotropy of $K_B = 4.29\ \mathrm{\frac{kerg}{cm^3}}$,  5.5 times stronger than the uniaxial term that is observed in CoFe(NH) structures.  In contrast, samples with silver and tungsten buffers present an isotropic angular dependence of both m$_r$/m$_s$ and H$_c$ parameters.

A notable enhancement of the uniaxial anisotropy is observed for CoFe\_Cu(H) structures while CoFe\_W(H) structures remains magnetically isotropic. The angular dependence of the remanence for the first system is well described by the superposition of uniaxial and biaxial anisotropies considering close values of the anisotropy constants for the two symmetries.

\begin{table}[ht]
\centering{}
\resizebox{\textwidth}{!}{
\begin{tabular}{|l|cccccccc|}
\hline
       \multicolumn{1}{|c|}{Sample}
       & \multicolumn{1}{c|}{$x_1$}    & \multicolumn{1}{c|}{$x_2$}   & \multicolumn{1}{c|}{$x_3$}    & \multicolumn{1}{c|}{$H^{EA}_{C}$ (Oe) }&\multicolumn{1}{c|}{$H^{HA}_{sat}\ (Oe) $}&\multicolumn{1}{c|}{$K_U\ (\mathrm{\frac{kerg}{cm^3}}) $}&\multicolumn{1}{c|}{$K_B\ (\mathrm{\frac{kerg}{cm^3}}) $}& \multicolumn{1}{c|}{$\gamma$}\\ \hline
CoFe\_Cu(NH) & \multicolumn{1}{c|}{0.20} & \multicolumn{1}{c|}{0.50}    & \multicolumn{1}{c|}{0.30}  &\multicolumn{1}{c|}{12.22} &\multicolumn{1}{c|}{18.96}   &\multicolumn{1}{c|}{0.75} &\multicolumn{1}{c|}{4.29}  &\multicolumn{1}{c|}{-}   \\ \hline
CoFe\_Ag(NH) & \multicolumn{3}{c|}{isotropic} &\multicolumn{1}{c|}{67.18}&\multicolumn{1}{c|}{504}  &\multicolumn{1}{c|}{-} &\multicolumn{1}{c|}{-} &\multicolumn{1}{c|}{-} \\ \hline
CoFe\_W(NH) & \multicolumn{3}{c|}{isotropic}&\multicolumn{1}{c|}{84.99}&\multicolumn{1}{c|}{82.78}&\multicolumn{1}{c|}{-}&\multicolumn{1}{c|}{-}&\multicolumn{1}{c|}{-}  \\ \hline
CoFe\_Cu(H)  & \multicolumn{1}{c|}{0.42} & \multicolumn{1}{c|}{0
43}    & \multicolumn{1}{c|}{0.15} &\multicolumn{1}{c|}{24.67}&\multicolumn{1}{c|}{46.57} &\multicolumn{1}{c|}{4.83}&\multicolumn{1}{c|}{5.03}&\multicolumn{1}{c|}{1.48}          \\ \hline
CoFe\_Ag(H)  & \multicolumn{1}{c|}{0.27} & \multicolumn{1}{c|}{0.64}    & \multicolumn{1}{c|}{0.08}  &\multicolumn{1}{c|}{67.01}  &\multicolumn{1}{c|}{709}&\multicolumn{1}{c|}{11.98} &\multicolumn{1}{c|}{14.88}  &\multicolumn{1}{c|}{0.59}    \\ \hline
CoFe\_W(H)  & \multicolumn{3}{c|}{isotropic}&\multicolumn{1}{c|}{67.07}&\multicolumn{1}{c|}{84.69} &\multicolumn{1}{c|}{-}&\multicolumn{1}{c|}{-} &\multicolumn{1}{c|}{-}     \\ \hline
\end{tabular}
}
\caption{\label{tabla:CoFe_mr}Parameters extracted from fitting the angular dependence of remanent magnetization and coercive field for the Co-Fe trilayers. The saturation field ($H^{HA}_{sat}$) and uniaxial anisotropy constants ($K_U$ and $K_B$) were obtained from hysteresis loops measured with the magnetic field applied along the hard and easy axes for the same set of samples.}
\end{table}

The angular dependence of the coercive field is shown in Fig. \ref{fig:CoFe-mr-hc}. For the copper-buffered CoFe films deposited without any applied magnetic field, in contrast with the remanence results, the coercivity exhibits an isotropic angular dependence. This kind of behavior falls outside the scope of the models presented in this work. The angular dependence of the coercive field in samples with copper and silver buffer layers deposited under an applied magnetic field is consistent with the remanence results. It is well described by a model assuming a superposition of a region with uniaxial anisotropy, which reverses following the two-phase model, and another region that reverses coherently under the influence of both biaxial and uniaxial anisotropies. A difference in the fitted $\gamma$ parameter is observed between copper- and silver-buffered samples. In both cases $\gamma < 2$, indicating that magnetization reversal occurs predominantly through domain rotation. However, the lower value obtained for the silver-buffered films ($\gamma = 0.59$) suggests a mechanism closer to the Stoner–Wohlfarth coherent rotation model, whereas the copper-buffered samples ($\gamma = 1.48$) deviate further from coherent behavior.

An alternative interpretation of the angular dependence observed in the polar diagrams is provided by the so-called hard-axis collapse model \cite{acosta2024assessment, geshev2021observation, idigoras2011collapse, idigoras2014magnetization, sedrpooshan2018magneto}. In this framework, the magnetic response is described in terms of two interacting single-domain grains with uniaxial anisotropy and misaligned easy axes, where the remanence along the hard axis arises from the relative orientation of the anisotropy axes and intergranular interactions. In contrast, in the present work we consider the coexistence of a uniaxial magnetocrystalline anisotropy with an additional biaxial anisotropy in certain regions of the sample. The fact that both models provide a good phenomenological description of the observed angular dependence of the remanence suggests that a distribution of uniaxial anisotropy axes combined with intergranular interactions can give rise to an effective biaxial anisotropy, consistent with the experimental observations.

Although the experimental results do not allow us to unambiguously determine whether the observed anisotropy is magnetocrystalline in origin or arises from a distribution of interacting grain easy axes, the buffer-induced structural characteristics may nevertheless offer insight into its possible origin. The anisotropies induced by the different buffer layers present a correlation with the grain structure of the films. Ni–Fe films deposited on Cu buffers exhibit a higher degree of structural order, and the presence of small epitaxial regions in these films cannot be excluded. Copper buffer layers are known to promote the growth of high-quality fcc metallic films \cite{chang1990magnetization, chang1990interface}, particularly for Fe–Ni alloys, due to the relatively small lattice mismatch ($\approx 2.58\%$). The presence of such local epitaxial regions may be responsible for the biaxial anisotropy observed in the magnetic measurements. In contrast, Ta buffers promote a (111) textured growth, which has been reported to favor the development of a uniaxial anisotropy and to reduce coercivity  \cite{MRSonline}. Finally, W buffers lead to a more disordered microstructure, as evidenced by the grain size analysis. As a consequence, a random distribution of magnetic easy axes is expected to dominate the magnetic behavior of these films.

The magnetic anisotropy observed in Co–Fe films is consistent with that reported for Fe–Ni systems. Films deposited on Cu buffers, which are expected to favor the formation of epitaxial regions, exhibit a biaxial magnetic anisotropy. In contrast, samples grown on W and Ag buffers display an isotropic magnetic response, which can be attributed to a random distribution of magnetic easy axes. In the case of Cu, the application of a magnetic field during growth leads to the coexistence of a buffer-induced biaxial anisotropy and a uniaxial anisotropy induced by the preferential alignment of Fe adatoms during deposition.

In spite of the fact Ag presents a relatively large lattice mismatch ($1\sim 5\%$), and epitaxial growth would not be expected a priori, the formation of semi-epitaxial films on Ag has been reported despite this mismatch. These films are characterized by strain that relaxes rapidly through the formation of nanometric crystallites, which grow coherently with the substrate rather than forming single-crystalline films  \cite{ASS05,goodall2001fabrication}. Our results suggest that the formation of such regions is enhanced by the presence of the applied magnetic field during growth. Co–Fe exhibits a significantly higher magnetocrystalline anisotropy and magnetostriction coefficient than Fe–Ni \cite{hall1959single}, which may explain why, in contrast to Fe–Ni, the coexistence of uniaxial and biaxial anisotropies is observed in Co–Fe films deposited under an applied magnetic field. Moreover, the biaxial anisotropy observed in the Ag-buffered sample is approximately three times larger than that measured for the Cu-buffered sample. Since higher strain is expected in films deposited on Ag due to lattice mismatch, the difference in the observed anisotropy can be attributed to magnetoelastic effects.

\section{Conclusions and discussion}

We investigated the influence of metallic buffer layers and field-assisted deposition on the magnetic anisotropy and magnetization reversal mechanisms of NiFe- and CoFe-based trilayers, with the aim of identifying configurations suitable for planar Hall effect (PHE) sensor applications. For NiFe films, we observe a clear correlation between the magnetic anisotropy and the degree of structural order of the grain ensemble. Films deposited on W buffers exhibit a broader grain-size distribution and an almost isotropic magnetic response, whereas films grown on Cu buffers show a higher degree of structural order and develop a biaxial magnetic anisotropy. We attribute these differences in grain-size distribution to the distinct melting points of the buffer layers, which affect the growth dynamics of the magnetic films. In addition, field-assisted deposition induces a dominant uniaxial anisotropy in NiFe films, with an anisotropy constant of approximately 3 $\frac{kerg}{cm^3}$, largely independent of the buffer layer.
In contrast, the magnetic response of CoFe films is primarily governed by the buffer layer. CoFe films grown on W buffers exhibit an almost isotropic magnetic behavior, consistent with a higher degree of structural disorder. For Cu buffers, a biaxial anisotropy of approximately 5  $\frac{kerg}{cm^3}$   is observed both with and without an applied magnetic field, indicating a dominant buffer-induced contribution. By contrast, Ag-buffered CoFe films exhibit a biaxial anisotropy only when deposited under an external magnetic field, reaching values up to 14.88 $\frac{kerg}{cm^3}$, suggesting that field-assisted deposition enhances the magnetic order promoted by the Ag buffer layer. The magnetization reversal in both systems can be successfully described in terms of an ensemble of regions with different anisotropy symmetries. Analysis of the angular dependence of the coercive field indicates that magnetization reversal is dominated by domain rotation for most samples, while only the NiFe film deposited on a Ta buffer under an applied magnetic field exhibits reversal governed by domain-wall motion. Based on these results, NiFe films grown on Ta buffer layers without an applied magnetic field emerge as the most suitable configuration for PHE sensor applications, as they combine low coercivity with a magnetization reversal driven by domain rotation, providing an optimal balance between sensitivity and stability.

\section{Acknowledgements}
The authors thank D. Mercado and A. Di Donato for technical assistance.
This work was supported by the National Agency for Scientific and
Technological Research (PICT-2021-CAT-I-00156) and the Argentinean
Ministry of Science, Technology and Innovation (PITES 7). 

\bibliographystyle{elsarticle-num}
\addcontentsline{toc}{section}{\refname}\bibliography{Biblio}

\appendix

\section*{\textemdash \textemdash \textemdash \textemdash \textemdash \textemdash \textemdash{}}
\end{document}